\documentclass[journal,12pt,draftclsnofoot,onecolumn]{IEEEtran}
\IEEEoverridecommandlockouts

\usepackage[utf8]{inputenc}
\usepackage[T1]{fontenc}
\usepackage{graphicx}
\graphicspath{{images/}}
\usepackage{amsmath,amssymb,amsfonts}
\usepackage{booktabs}
\usepackage{algorithm}
\usepackage{algpseudocode}
\usepackage[hidelinks]{hyperref}

\newcommand{\E}{\mathbb{E}}
\newcommand{\ipr}{\nu^2}            % inverse participation ratio
\newcommand{\lcrit}{\ell^{*}}       % critical sampling exponent
\usepackage{xcolor}

\begin{document}

\title{Adaptive Row Selection Meets Asynchrony in Randomized Kaczmarz}

\author{\IEEEauthorblockN{Evan Coleman}
\thanks{\copyright~2026 IEEE. Personal use of this material is
    permitted. Permission from IEEE must be obtained for all other
    uses, in any current or future media, including
    reprinting/republishing this material for advertising or
    promotional purposes, creating new collective works, for resale or
    redistribution to servers or lists, or reuse of any copyrighted
    component of this work in other works. Accepted for publication in
    the Proceedings of the 2026 IEEE High Performance Extreme Computing
    Conference (HPEC).}
\IEEEauthorblockA{Department of Computer Science\\
University of Mary Washington\\
Fredericksburg, Virginia, USA\\
ecolema4@umw.edu}}

\maketitle

% ===========================================================================
\begin{abstract}
Randomized Kaczmarz is a natural fit for large sparse least-squares and tomographic reconstruction, and adaptive row selection can reduce iteration counts. However, deploying adaptive selection on a shared-memory machine means sampling from a residual that lock-free workers are concurrently modifying, often using stale data. We present the first systematic study of this regime: residual-weighted and greedy Kaczmarz under asynchronous execution, measured across 339 runs on a 96-core node with realized (not injected) delays. Four findings carry directly to practice. (i) Stability is governed by a boundary $\ell^*(T)$ between sampling aggressiveness and thread count; below it, more aggressive sampling is strictly better, so one should tune to just inside the cliff. (ii) Threshold-greedy selection (the standard accelerated rule) is unstable at high thread counts, diverging almost immediately. (iii) Under-relaxation buys back the cliff at a predictable cost, giving a usable safety knob. (iv) Consistent-snapshot reads admit a rare, scheduling-dependent divergence that live (inconsistent) reads never exhibited; live reads also cost less per step, making them the right default. We validate the implementation against published sequential results and outline the distributed two-level sampler these measurements motivate.
\end{abstract}


% ===========================================================================
\section{Introduction}\label{sec:intro}
Randomized Kaczmarz~\cite{strohmer2009randomized} is a row-action solver: each step projects the current iterate onto one equation's hyperplane, costing $O(\mathrm{nnz})$ of a single row with no factorization. That structure makes it a natural choice for large sparse least-squares and for applications such as tomographic reconstruction, where it is the algebraic backbone of the classical Algebraic Reconstruction Technique (ART) method. \emph{Adaptive selection} (e.g., greedy~\cite{bai2018greedy} or residual-power-weighted~\cite{steinerberger2021weighted,haddock2021greed,gower2021adaptive}) accelerates convergence by preferring high-residual rows; \emph{asynchronous execution}~\cite{avron2015revisiting,liu2014asynchronous} extracts parallelism by letting lock-free workers update shared state without barriers. 

Each of these mechanisms is well understood in isolation, but deploying them together creates a situation neither analysis covers. Adaptive selection must rank rows by residual magnitude; but on an asynchronous machine the residual is a shared vector that other workers are concurrently modifying (possibly with stale views). The selector is therefore choosing from an imperfect distribution, where the discrepancy from ideal grows with the number of in-flight updates. Existing adaptive-Kaczmarz theory assumes the residual is current; existing asynchronous theory assumes the sampling distribution is fixed. What actually happens between those assumptions has not been measured, and existing parallel implementations~\cite{wang2022prkp,bolukbacsi2024distributed,panchal2026rgdbek} do not analyze it.

Using the power-weighted family $P(i)\propto|r_i|^\ell$ (a single knob from uniform, $\ell\to 0$, through greedy, $\ell\to\infty$) we instrument residual-weighted and threshold-greedy Kaczmarz across 339 runs on a 96-core node  with delays that are realized by the hardware rather than synthetically injected, on: (1) a benign dense problem, (2) a coherent tomographic problem, and (3) the standard sparse suite. We find that the read-consistency effect known from asynchronous Jacobi does not govern here \cite{coleman2026rwrj}. What governs is a stability boundary between how aggressively selection chases residual peaks and how many workers are running, a cliff that aggressive sampling races toward as concurrency grows. The contributions of this paper are:

\begin{itemize}
    \item \textbf{A stability boundary $\lcrit(T)$:} We map a boundary in the (sampling exponent $\ell$, thread count $T$) plane that decreases in $T$: more concurrency tolerates less aggressive sampling. Below the boundary greedier sampling is strictly faster so tuning amounts to a race to a concurrency-dependent cliff.
    \item \textbf{Greedy selection is asynchrony-incompatible at scale:} Threshold-greedy selection~\cite{bai2018greedy}, the standard accelerated rule, sits outside the boundary at high thread count: it diverges within a fraction of a sweep at $T=96$ on every test problem, under either read semantics.
    \item \textbf{Step-size control recovers the cliff:} Under-relaxation restores convergence past the boundary in our experiments: every configuration that diverges at full step ($\beta=1$) converges at $\beta\le 0.5$ (21 of 21 runs), at a rate cost of $\beta(2-\beta)$ times a measured interference premium of roughly two.
    \item \textbf{Read-mode semantics as tail risk:} Consistent-snapshot reads admit a rare, scheduling-dependent divergence (one of 36 runs at a coherent cell; the same-seed survivor differs only in thread interleaving) that inconsistent (live) reads never exhibited (zero of 36). Inconsistent reads are also cheaper per step, which makes them the right default.
    \item \textbf{A validated, exact implementation:} Sequential greedy/uniform speedups reproduce published results~\cite{bai2018greedy} where the row-normalization protocols coincide, and the maintained residual agrees with a recomputed $b-Ax$ at every exit (to within $2.4\times10^{-12}$ on converged runs and to within $3.8\times10^{-15}$ in relative terms at the divergence flag), so every reported divergence is numerical behavior, not a data race.
\end{itemize}
% ===========================================================================
\section{Related Work}\label{sec:related}

The Kaczmarz method projects the iterate onto one equation's hyperplane per step. Its tomography connection is classical: Algebraic Reconstruction Technique
(ART) applies this row-action projection view to tomographic projection
equations~\cite{gordon1970algebraic}. Strohmer and Vershynin~\cite{strohmer2009randomized} showed that sampling row $i$ with probability proportional to $\|a_i\|^2$ yields expected linear convergence at a rate set by the scaled condition number; Leventhal and Lewis~\cite{leventhal2010randomized} cast this as randomized coordinate descent on the normal equations, which is the view we use here. Adaptive rules sharpen the constant by steering selection toward large residuals. Bai and Wu~\cite{bai2018greedy} introduced greedy randomized Kaczmarz (GRK), which thresholds the residual and samples among the rows above the cutoff; Steinerberger~\cite{steinerberger2021weighted} analyzed the continuous family $P(i)\propto|r_i|^p$, recovering uniform sampling at $p=0$ and the maximal-residual Motzkin rule as $p\to\infty$, with the $p=1$ case appearing in adaptive sketch-and-project~\cite{gower2021adaptive} and the same power family extended to sparse recovery in a Bregman setting~\cite{zhang2023weighted}. The Motzkin and sampling-Kaczmarz–Motzkin line~\cite{nutini2016convergence,haddock2019motzkin,haddock2021greed} quantifies the acceleration through the residual's dynamic range. Across all of this work, selection is computed from the \emph{current} residual (a synchronous assumption). Parallelism and hardware optimization, where considered, typically remain synchronous: averaging independent projections~\cite{moorman2021randomized} changes the iteration, while randomized block Kaczmarz~\cite{needell2015randomized} processes multiple rows simultaneously to increase arithmetic intensity and mitigate the memory-bound nature of single-row updates.

Asynchronous linear solvers trace to chaotic relaxation~\cite{chazan1969chaotic} and were given a modern asynchronous \emph{and} randomized analysis by Avron, Druinsky, and Gupta~\cite{avron2015revisiting}, who proved convergence of shared-memory asynchronous updates under bounded staleness. More broadly, lock-free shared-memory stochastic updates were popularized by HOGWILD!~\cite{recht2011hogwild}, where sparsity makes overwrites tolerable. For Kaczmarz specifically, Liu, Wright, and Sridhar~\cite{liu2014asynchronous} analyzed an asynchronous parallel variant, but under two assumptions we revisit: a \emph{fixed} (non-adaptive) sampling distribution, and (effectively) consistent reads, argued to be benign because each update touches only one row's sparse support. Component-averaged row projection (CARP)~\cite{gordon2005carp} parallelizes row actions by a different route. Most relevant are the recent parallel and distributed implementations
that pair residual-adaptive selection with parallel execution.
PRKP~\cite{wang2022prkp} combines greedy row sampling with lazy
residual estimates on distributed memory, and the distributed-memory
solver of~\cite{bolukbacsi2024distributed} partitions rows across
ranks. In contrast, RGDBEK~\cite{panchal2026rgdbek} parallelizes
residual-derived block selection within a strictly bulk-synchronous
MPI+GPU design that avoids staleness by construction. None of these
works analyzes adaptive selection on stale residual
information (the synchronous designs do not encounter it), so the
regime where adaptive sampling and asynchrony interact remains
unmeasured.

% ===========================================================================
\section{Algorithm and Implementation}\label{sec:method}
\subsection{Power-weighted sampling}
We solve a consistent $m \times n$ system $Ax=b$, with rows normalized so
$\|a_i\|=1$ and $b$ scaled accordingly, following the
Steinerberger/AsyRK convention~\cite{steinerberger2021weighted,liu2014asynchronous}.
A relaxed Kaczmarz step on row $i$ is
\[
    x \leftarrow x + \beta r_i a_i,\qquad r=b-Ax ,
\]
so $\beta=1$ exactly satisfies equation $i$. Rows are sampled as
$P(i) \propto |r_i|^\ell$.
The exponent $\ell$ interpolates from uniform sampling
($\ell=0$, equivalent to Strohmer--Vershynin for unit rows) to
the maximal-residual Motzkin rule ($\ell\to\infty$). We also
test threshold-greedy GRK~\cite{bai2018greedy}, and a deterministic cyclic baseline that visits rows in fixed order. To summarize
how concentrated the residual is, we track its inverse participation ratio (IPR)
\[
    \nu^2(r)=m\|r\|_4^4/\|r\|_2^4\in[1,m],
\]
which gives the effective number of large residual entries.

\subsection{Residual maintenance}
Rather than recompute $r=b-Ax$, workers maintain it exactly
via the Gram-row update of \cite{steinerberger2021weighted}.
Committing $\lambda=\beta r_i^{\rm used}$ on row $i$ updates
\[
    x \leftarrow x+\lambda a_i,\qquad
    r \leftarrow r-\lambda Q_{i,:},
\]
where $Q=AA^\top$ and $Q_{ii}=1$. Thus row $i$ of $Q$ is the
set of equations coupled to row $i$ through overlapping support,
analogous to a stencil row in Jacobi. For this measurement study
$Q$ is precomputed once in sparse form. Each update costs
$O(m)$ for inverse-CDF sampling plus $O(\deg_Q(i))$ for the
commit; the sampling pass dominates. As a check, every run joins its 
workers at termination, recomputes 
$b-Ax$, and compares it to the maintained residual.

\subsection{Read semantics}
Workers are lock-free, so a worker may sample from a stale
residual. We compare two policies. With \emph{consistent} reads,
the worker snapshots $r$ once and uses that snapshot both to
sample $i$ and to form $r_i^{\rm used}$. With \emph{inconsistent}
reads, the sampling CDF is built from live atomic reads and
$r_i$ is read again after $i$ is selected. Thus the value used for
the update may differ from the value that caused the row to be
selected. Consistent reads pay an $O(m)$ copy per step; GRK
adds another source of staleness because its threshold and
sampling passes may see different live residuals.

\begin{algorithm}[t]
\caption{Per-worker loop (power-weighted, exponent $\ell$, step $\beta$)}
\label{alg:worker}
\begin{algorithmic}[1]
\While{not converged}
  \State $d \gets$ atomic load of global commit counter \Comment{dispatch}
  \State obtain residual view of $r$ \Comment{snapshot or live reads}
  \State sample row $i$ with $P(i) \propto |r_i|^{\ell}$ \Comment{inverse CDF}
  \State $r_i^{\mathrm{used}} \gets$ \textbf{if} inconsistent \textbf{then} re-read $r_i$ \textbf{else} snapshot $r_i$
  \State $\lambda \gets \beta\, r_i^{\mathrm{used}}$
  \State atomic $x \mathrel{+}= \lambda\, a_i$ \Comment{support of row $i$}
  \State atomic $r \mathrel{-}= \lambda\, Q_{i,:}$ \Comment{Gram row $=$ stencil}
  \State $c \gets$ fetch-and-add commit counter; record delay $c - d$
\EndWhile
\end{algorithmic}
\end{algorithm}

% ===========================================================================
\section{Experimental Setup}\label{sec:setup}

All experiments ran on a 96-core ACES node
(Texas A\&M), with threads pinned using
\texttt{OMP\_PROC\_BIND=spread} and \texttt{OMP\_PLACES=cores}.
Staleness was measured, not injected: each worker records the global
commit counter when it begins a step and again when it commits, and
we report the realized dispatch-to-commit gap. Thus the delay
distribution is a property of the hardware and schedule. All test
systems are consistent by construction: we fix $x^\star$ and set
$b=Ax^\star$, so both the relative residual $\|r\|/\|b\|$ and the
relative solution error $\|x-x^\star\|_2/\|x^\star\|_2$ are available.
Runs stop at $\|r\|/\|b\|<10^{-6}$, are flagged divergent when
$\|r\|/\|b\|>10^8$, and are capped by a problem-dependent sweep
limit; one sweep denotes $m$ commits, a full pass over the rows in expectation. Paired read-mode comparisons use matched seeds, so each
consistent/inconsistent pair shares the matrix, right-hand side, and
per-thread random streams and differs only in read policy. Table~\ref{tab:problems}
summarizes the test problems: a benign dense system, coherent
tomographic systems at two scales, and 16 SuiteSparse matrices from
the Bai--Wu test suite~\cite{bai2018greedy}. 
The tomographic systems use a parallel-beam geometry with 90 angles
$\times$ 93 detector bins on a $64\times64$ grid (tomo-4x:
$180\times185$ on $128\times128$); rays that miss the grid are dropped.
In total we report 339
instrumented runs. At termination, every run joins its workers, recomputes $b-Ax$, and
compares it to the maintained residual; the maximum discrepancy across
converged runs is $2.4\times10^{-12}$.

\begin{table}[t]
    \centering
    \setlength{\tabcolsep}{3.5pt}
    \caption{Test problems and measured static structure (empty rows dropped, rows normalized, $b=Ax^\star$). $\overline{\deg}_Q$ is the mean Gram row degree (coupling density); $\bar\rho_2^{\,\mathrm{off}}$ the mean off-diagonal Gram row 2-norm; $\bar\chi=\E|Q_{ij}|$ the mean pairwise coupling; $\ipr(b)$ the initial residual concentration. The SuiteSparse validation matrices span $m=10$--$2063$, $n=10$--$24310$, $\overline{\deg}_Q=4.76$--$400$, and $10^3\bar\chi=1.8$--$768$.}
    \label{tab:problems}
    \small
    \begin{tabular}{@{}lrrrrr@{}}
        \toprule
        Problem & $m \times n$ & $\overline{\deg}_Q$ & $\bar\rho_2^{\,\mathrm{off}}$ & $10^{3}\bar\chi$ & $\ipr(b)$ \\
        \midrule
        gaussian     & 4{,}000 $\times$  1{,}000  & 4{,}000  & 2.00 & 25.2 & 2.9 \\
        gaussian-4x  & 16{,}000 $\times$ 1{,}000  & 16{,}000 & 4.00 & 25.2 & 2.9 \\
        tomo         & 7{,}546 $\times$ 4{,}096  & 3{,}016  & 2.58 & 12.0 & 4.5--6.7 \\
        tomo (point) & 7{,}546 $\times$ 4{,}096  & 3{,}016  & 2.58 & 12.0 & 47.0 \\
        tomo-4x      & 29{,}980 $\times$ 16{,}384 & 11{,}866 & 2.82 & 6.0  & 5.3 \\
        \bottomrule
    \end{tabular}
    \\[2pt]
    {\footnotesize At $T{=}96$: $\tau\bar\chi = 2.40$ (gaussian), 1.14 (tomo), 0.57 (tomo-4x).}
\end{table}

% ===========================================================================
\section{Results}\label{sec:results}

\subsection{Sampling aggressiveness under asynchrony}\label{sec:results:samplers}

\begin{figure*}[t]
\centering
\includegraphics[width=0.95\textwidth]{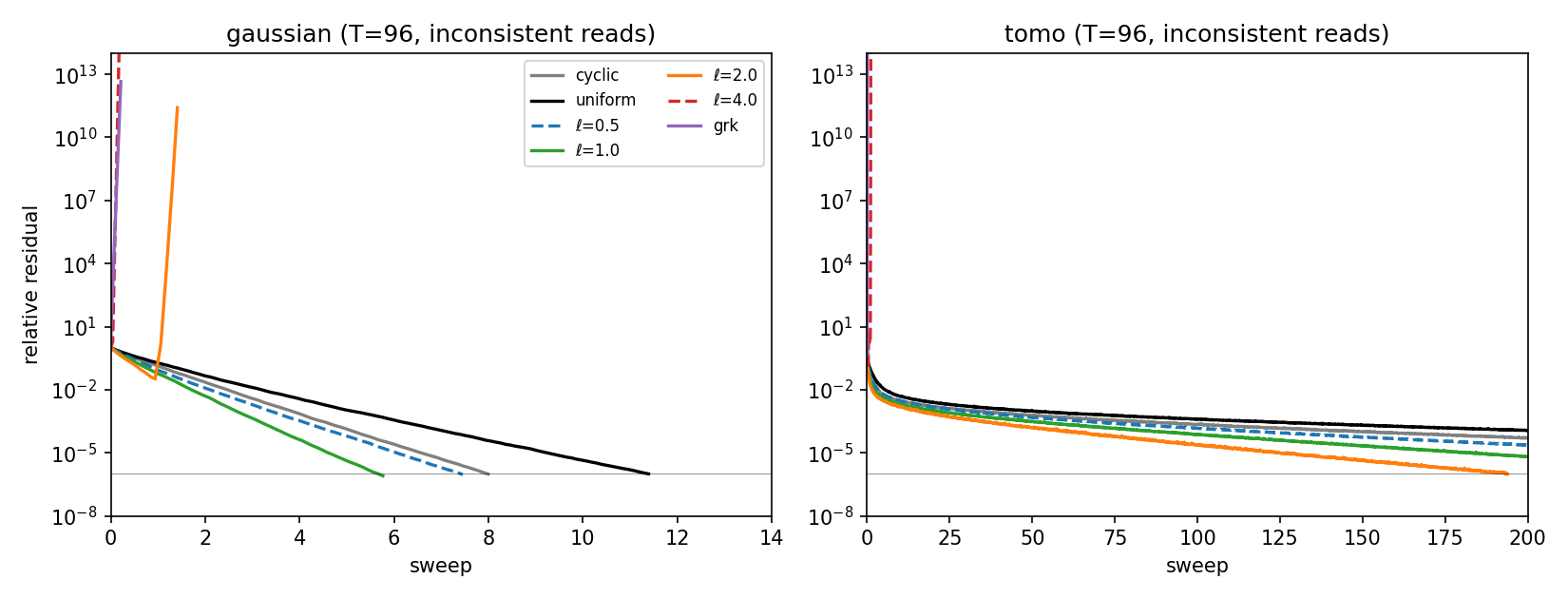}
\caption{Relative-residual trajectories by sampling rule at $T=96$
under inconsistent reads. On gaussian, $\ell=1$ is the fastest
survivor while $\ell\ge2$ and GRK diverge. On tomo, the stability
cliff occurs at a larger exponent, but GRK again fails rapidly.
The optimum is therefore interior: residual bias helps until
asynchrony drives the sampler past a problem-dependent cliff.}
\label{fig:samplers}
\end{figure*}

We begin with the observation that motivates the rest of the
paper. Figure~\ref{fig:samplers} compares cyclic, uniform,
power-weighted sampling with $\ell\in\{0.5,1,2,4\}$, and
threshold-greedy GRK at $T=96$ under inconsistent reads on the
two built-in problems. Synchronous theory suggests that stronger
residual bias should reduce iteration counts. Under asynchrony,
that ordering holds only up to a stability cliff.
On the gaussian problem, cyclic, uniform, and $\ell\le1$ converge;
$\ell=1$ reaches tolerance in 5.8 sweeps, roughly twice as fast as
uniform's 11.4 sweeps. In contrast, $\ell\ge2$ diverges. GRK is the
extreme case: on both problems and under either read policy, it
fails within a fraction of a sweep, reaching residuals of
$10^{12}$--$10^{21}$. Its threshold rule concentrates selection on
the largest residual entries, so many workers repeatedly update the
same coupled neighborhoods.

The $\ell=2$ gaussian run shows the mechanism more clearly because
it first behaves like a successful solve. The residual falls to about
$3\times10^{-2}$ within one sweep while the concentration statistic
$\nu^2$ remains in the range 3--6; then the run diverges within the
next half-sweep as $\nu^2$ rises to 130--170. We interpret this as
an interference floor. While the true residual is large, it dominates
the perturbations caused by the $\Theta(T)$ in-flight updates. Once
the residual reaches the same scale as those perturbations, a
residual-weighted sampler begins ranking interference rather than
signal. The selected neighborhoods are then over-updated, the
perturbation grows, and the feedback loop closes.
Uniform and cyclic sampling do not rank residual entries, so they
do not amplify this floor in the same way. This first experiment
therefore points to sampling aggressiveness, rather than read
consistency alone, as the axis that must be mapped next.

\subsection{The stability boundary $\lcrit(T)$}\label{sec:results:boundary}

\begin{figure}[t]
\centering
\includegraphics[width=\columnwidth]{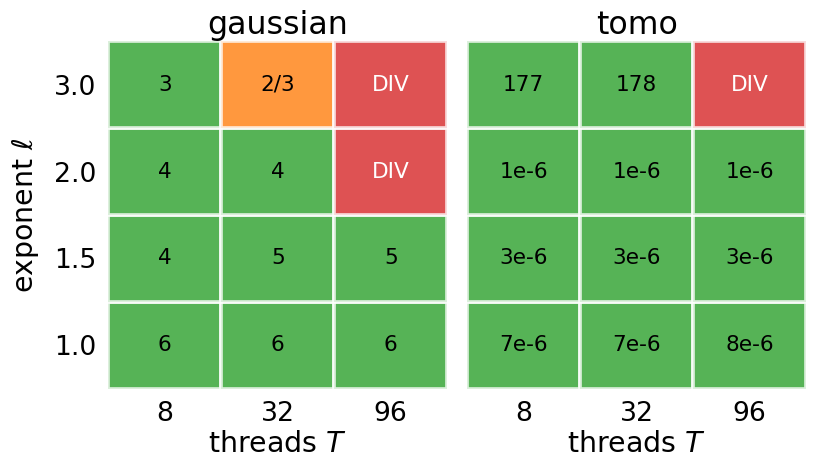}
\caption{Stability boundary $\lcrit(T)$ over the (sampling exponent, thread count) plane; three seeds per cell, green annotated with median sweeps to tolerance (or relative residual at the 200-sweep cap). The boundary decreases with concurrency, and below it the synchronous ordering survives (greedier is strictly faster) so tuning is a race to a concurrency-dependent cliff. The mixed cell (two converge, one diverges) shows the boundary is probabilistic.}
\label{fig:map}
\end{figure}

Figure~\ref{fig:map} maps the outcome over the
$(\ell,T)$ plane, with three seeds per cell. A
concurrency-dependent boundary $\lcrit(T)$ separates convergence from
divergence, and it moves downward as $T$ grows: more workers
tolerate less aggressive residual bias. On gaussian, $\ell=3$
is stable at $T=8$, mixed at $T=32$ (two seeds converge and
one diverges), and unstable at $T=96$; the high-concurrency
cliff lies between $\ell=1.5$ and $\ell=2$. On tomo, the
boundary is shifted upward: $\ell=3$ remains stable through
$T=32$ and fails only at $T=96$. In the tomo panel of
Fig.~\ref{fig:map}, $\ell\le2$ cells report the residual at the
200-sweep cap, essentially at tolerance; the $\ell=3$, $T\le32$ cells
converge outright in $\sim$177 sweeps.

Below the boundary, the synchronous ordering largely survives.
On gaussian, median sweep counts decrease as $\ell$ increases
within the stable region, so the best setting is not uniform
sampling but the largest exponent still inside the cliff. Thus
tuning becomes a race to a problem- and concurrency-dependent
boundary.

The static statistics in Table~\ref{tab:problems} indicate what
moves that boundary. The normalized off-diagonal Gram-row norm
$\bar\rho^{\rm off}_2/\sqrt m$ is nearly the same for gaussian
and tomo and is $m$-independent for gaussian, yet their cliffs
differ by about one unit of $\ell$. The better discriminator is
the mean pairwise coupling $\bar\chi=E|Q_{ij}|$: 
$10^3\bar\chi=25.2$ for gaussian, $12.0$ for tomo, and $6.0$
for tomo-4x, exactly matching the observed stability ordering.
With mean realized delay $\tau\approx T-1$, the corresponding
interference scale $\tau\bar\chi$ at $T=96$ is $2.40$, $1.14$,
and $0.57$; Fig.~\ref{fig:predictor} tests this discriminator
suite-wide.

This also explains why gaussian-4x rescues $\ell=2$ at
$T=96$, changing from 0/6 converged runs to 6/6. Its pairwise
coupling is the same as gaussian, so the interference scale has
not changed. What changes is the convergence clock: with
$m/n=16$, the system reaches tolerance in roughly one sweep,
before the interference floor that destroys the smaller gaussian
instance can dominate. The stability cliff is therefore a race
between convergence and interference growth.

\subsection{Step-size control at the cliff}

\begin{figure*}[t]
\centering
\includegraphics[width=0.95\textwidth]{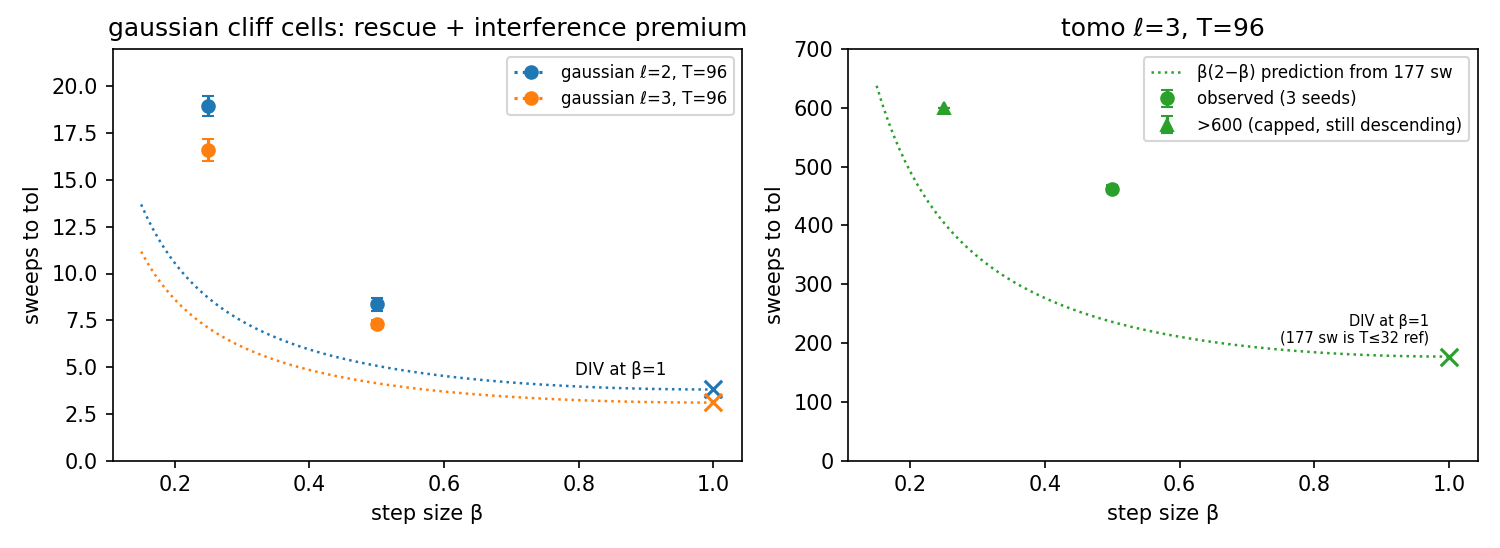}
\caption{Sweeps to tolerance versus relaxation parameter $\beta$ at
cliff cells that diverge for $\beta=1$. Dotted curves show the
synchronous $1/[\beta(2-\beta)]$ prediction. Under-relaxation restores
convergence in all tested rescue runs, but with a $1.7$--$2.3\times$
asynchronous premium beyond the synchronous prediction.}
\label{fig:beta}
\end{figure*}

If instability occurs because workers over-commit to an
interference-contaminated residual, then reducing the projection
step should improve stability. Figure~\ref{fig:beta} reruns
the cliff cells from Figure~\ref{fig:map} with smaller
relaxation parameter $\beta$. Every configuration that diverges at
full step, $\beta=1$, converges for some $\beta\le 0.5$; across the
21 tested rescue runs, none diverged. The metastable gaussian cell
at $(\ell=3,T=32)$ also improves from two-of-three to three-of-three
converged seeds.

The recovery is not free. In the synchronous Kaczmarz iteration,
under-relaxation predicts a sweep-count penalty proportional to
$1/[\beta(2-\beta)]$. The observed penalty is larger by about
$1.7$--$2.3\times$, with the excess growing as $\beta$ decreases.
Thus $\beta$ damps the instability, but it also slows the correction
of the rows that the stale sampler continues to over-select. This
extra cost is the asynchronous interference premium.

The engineering conclusion is that $\beta$ is a reliable safety
knob: when concurrency pushes the desired sampling exponent beyond
the stability boundary, under-relaxation can buy back convergence at
a measurable rate cost. The tomo $\beta=0.25$ runs are censored at
the 600-sweep cap but continue descending through
$1.3$--$1.5\times10^{-5}$, and are marked accordingly in
Figure~\ref{fig:beta}.

\begin{figure*}[t!]
\centering
\includegraphics[width=0.95\textwidth]{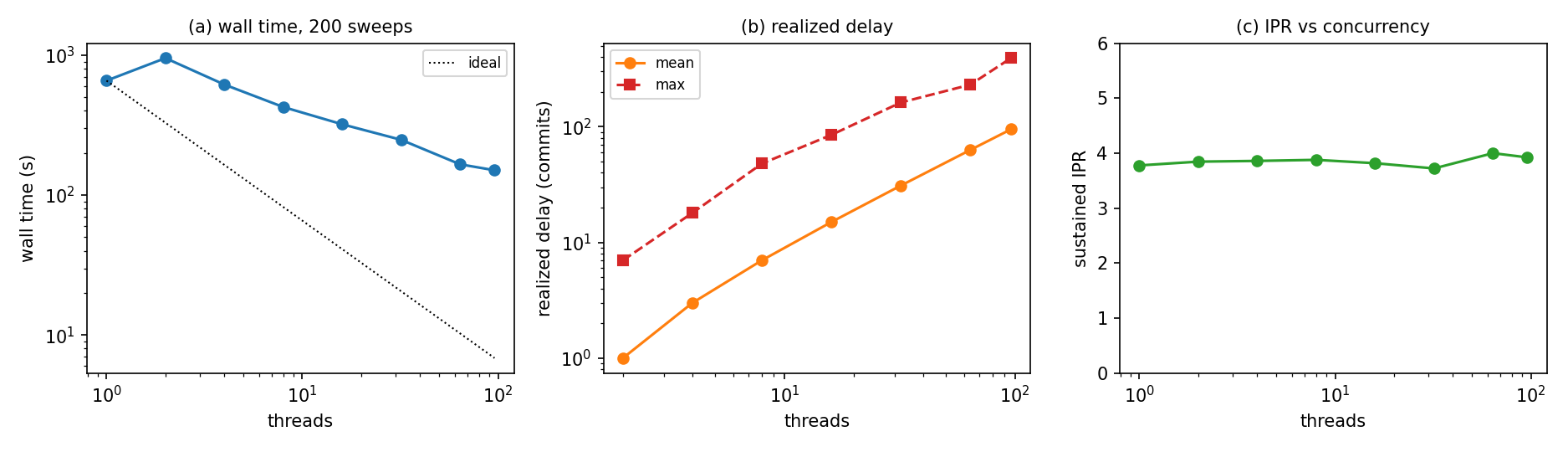}
\caption{Scaling on the headline tomo instance. Wall-clock speedup
saturates at $4.4\times$ because the $O(m)$ global sampling pass is
memory-bound. In contrast, sweeps to tolerance and sustained
$\nu^2$ remain nearly flat through $T=96$, so asynchrony is cheap in
iteration terms but not in wall-clock time.}
\label{fig:scaling}
\end{figure*}

\subsection{Read-mode semantics as tail risk}

Read consistency does not set the stability boundary, but it
does affect the tail behavior. In the paired tomo runs at
$T=96,\ell=2$, de-censored to 400 sweeps, one consistent-read
run diverges catastrophically at sweep 2.3
($\|r\|/\|b\|\approx5\times10^{27}$), while its matched
inconsistent-read twin converges. The pair uses the same matrix,
right-hand side, and per-thread random streams; it differs only in
read policy and thread interleaving. Across all matched tomo runs,
consistent reads had 1 catastrophic divergence in 36 trials, while
inconsistent reads had 0 in 36 (Table~\ref{tab:tail}). The confidence
intervals overlap,
so we do not estimate a failure rate. The result is instead an
existence proof: snapshot reads admit a scheduling-dependent
early-divergence mode that live reads did not exhibit in our runs.

This observation also favors live reads on cost. Consistent reads
copy the full residual once per step, whereas live reads build the
sampling CDF directly from atomic loads and re-read the selected
entry before committing. Since live reads were cheaper and showed
no tail failures in these matched tests, they are the better default
for the broad-coupling regime studied here.

Finally, we tested whether the collision-dominated Jacobi regime
appears in tomography by using a point phantom with much higher
initial concentration, $\nu^2(b)\approx47$. Both read policies
converged in all six runs, with similar sweep counts. The protection
is structural: a point object's sinogram is supported on only a few
rays per angle but across \emph{all} angles, so the Radon transform
caps achievable concentration at roughly $m/n_{\mathrm{angles}}$
($\approx84$ here, against the measured 47) --- orders of magnitude
below the $\ipr\sim m$ that same-row pile-up requires. Thus, for these
Kaczmarz problems, the main failure mode is not read inconsistency
itself but residual-adaptive sampling of an interference floor.

\begin{table}[t]
\centering
\caption{Catastrophic divergences at the coherent cell (tomo, $T{=}96$,
$\ell{=}2$), with exact 95\% Clopper--Pearson intervals. The intervals
overlap: the claim is existence and asymmetry, not a rate difference.}
\label{tab:tail}
\small
\begin{tabular}{@{}lccc@{}}
\toprule
read mode & catastrophic / runs & rate & 95\% CI \\
\midrule
consistent   & 1 / 36 & 2.8\% & $[0.1\%,\ 14.5\%]$ \\
inconsistent & 0 / 36 & 0\%   & $[0\%,\ 9.7\%]$ \\
\bottomrule
\end{tabular}
\end{table}

\begin{figure}[t]
\centering
\includegraphics[width=\columnwidth]{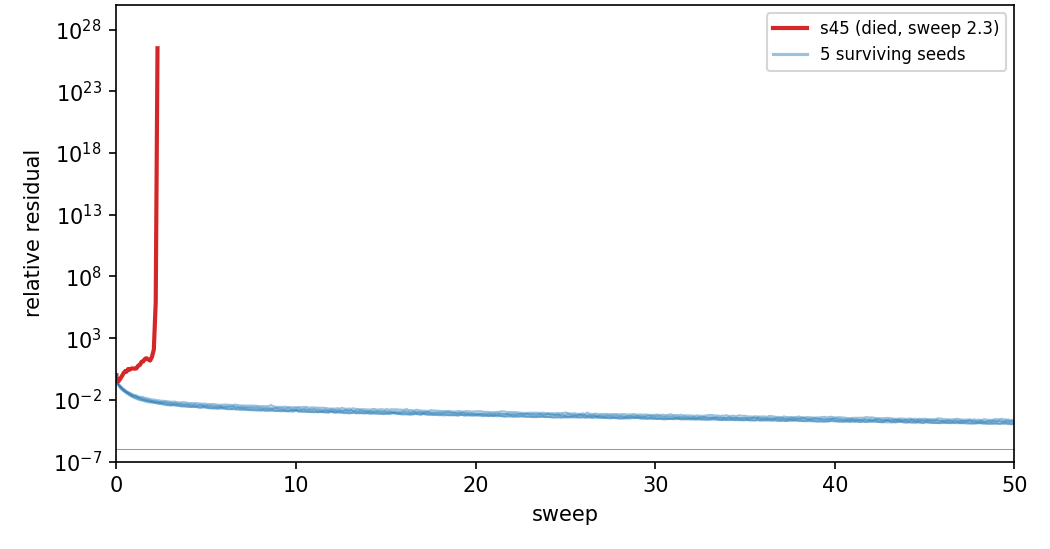}
\caption{Six same-configuration consistent-reads runs, tomo, $T{=}96$, $\ell{=}2$. Five converge; one diverges catastrophically at sweep 2.3 (residual $5\times10^{27}$), while its same-seed twin survives; the runs differ only in thread interleaving, so the failure is scheduling-dependent, not deterministic.}
\label{fig:tail}
\end{figure}

\subsection{Implementation soundness and scaling}\label{sec:results:scaling}

The divergences above are properties of the asynchronous iteration,
not residual-maintenance errors. 
On every converged run, the maintained residual agrees with a recomputed
$b-Ax$ to within $2.4\times10^{-12}$; and in validation re-runs of every
divergent $(\ell,T)$ cell and every greedy configuration (33 runs spanning
both read modes and all five test problems), the maintained and recomputed
residuals at the divergence flag (where $\|r\|/\|b\|$ reaches
$10^{8}$--$10^{82}$) agree to within $3.8\times10^{-15}$ in relative
terms: the explosive trajectories are computed exactly.
The sequential implementation also reproduces
published GRK-over-uniform iteration-count ratios~\cite{bai2018greedy} on the
near-constant-row-norm matrices where the normalization protocols
coincide: bibd gives ratios $2.9/3.0/3.0$ versus their
$3.2/3.1/3.3$, df2177 gives $4.9$ versus $5.0$, crew1 gives
$5.0$ versus $4.5$, WorldCities gives $6.1$ versus $5.7$, and
cari gives $4.2$ versus $4.7$. The main disagreements occur on
heterogeneous-norm or ill-conditioned matrices, where row
normalization changes the effective problem and residual stopping
need not track solution error.

Figure~\ref{fig:scaling} separates iteration behavior from wall-clock
behavior on the headline tomo instance. Sweeps to tolerance are nearly
flat from $T=1$ to $T=96$, and the sustained concentration statistic
remains $\nu^2\approx3.7$--$4.0$, so asynchrony is essentially free
in iteration terms. Wall-clock scaling is much weaker: speedup reaches
only $4.4\times$ on 96 threads, and $T=2$ is slower than $T=1$.
The bottleneck is the $O(m)$ atomic-read pass used to build the global
sampling CDF, which is memory-bound and dominates the sparse
Gram-row commit.

\begin{figure}[t]
\centering
\includegraphics[width=\columnwidth]{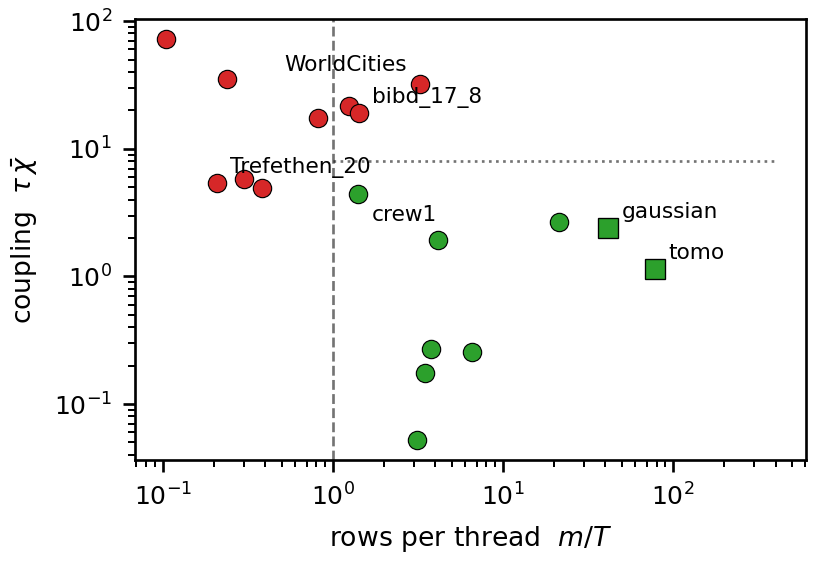}
\caption{Asynchronous survival predicted from two static quantities
(uniform sampling, $T{=}96$): rows-per-thread $m/T$ vs.\ in-flight
coupling $\tau\bar\chi$; green converged, red diverged. crew1 and
bibd\_17\_8 share $m/T{=}1.4$ but differ in coupling and fate.}
\label{fig:predictor}
\end{figure}

Rows per thread and coupling are therefore joint stability
resources, and both are computable from $A$ before any run.
Figure~\ref{fig:predictor} plots the 16 SuiteSparse matrices in the
$(m/T,\ \tau\bar\chi)$ plane against their $T=96$ uniform-sampling
outcomes. Below roughly one row per thread nothing survives,
regardless of coupling; above that floor, survival tracks
$\tau\bar\chi$. The sharpest evidence is a natural controlled pair:
crew1 and bibd\_17\_8 sit at identical $m/T=1.4$ with opposite
outcomes, separated only by coupling ($\tau\bar\chi$ of $4.4$ versus
$19.0$); WorldCities, the lone high-$m/T$ casualty, carries the
largest coupling in the suite ($\tau\bar\chi=32$).

% ===========================================================================
\section{Conclusion}\label{sec:conclusion}
Asynchronous Kaczmarz with residual-adaptive selection is governed by coupling structure. Unlike the narrow-coupling, collision-bound regime of asynchronous Jacobi~\cite{coleman2026rwrj}, the broad-coupling regime of Kaczmarz (e.g., dense least-squares and tomography) is bound by the product of sampling aggressiveness and coupling density, which creates a neighborhood interference floor. Measurements identify the discriminator concretely: the mean pairwise coupling $\bar\chi=\E|Q_{ij}|$ strictly orders the stability boundary, whereas spectral proxies ($\bar\rho_2^{\mathrm{off}}/\sqrt m$) and worst-case coherence fail to predict it.

The results here suggest it would be optimal to default to inconsistent reads, which showed no tail divergences in our runs and cost less per step; tune the sampling exponent just inside the boundary $\lcrit(T)$ where aggressiveness is strictly beneficial; recover the cliff with under-relaxation if concurrency forces a smaller $\ell$ than desired; and budget at least $O(10)$ rows per thread. 

Two directions seem ideal for further exploration: (1) developing the staleness-aware theory behind the $\lcrit(T)$ boundary and super-quadratic under-relaxation premium, and (2) the observed $O(m)$ global sampling bottleneck motivates work on a distributed two-level sampler, where block-level masses refreshed by asynchronous reductions explicitly decouple sampling staleness from iterate staleness.

\section*{Acknowledgment}
This work leveraged the ACES cluster at Texas A\&M University under
allocation CIS250436 from the ACCESS program (NSF grants \#2138259,
\#2138286, \#2138307, \#2137603, \#2138296).

% ===========================================================================
\bibliographystyle{IEEEtran}
\bibliography{refs}

@article{avron2015revisiting,
  author  = {Avron, Haim and Druinsky, Alex and Gupta, Anshul},
  title   = {Revisiting Asynchronous Linear Solvers: Provable Convergence
             Rate Through Randomization},
  journal = {Journal of the ACM},
  volume  = {62},
  number  = {6},
  pages   = {51:1--51:27},
  year    = {2015},
  doi     = {10.1145/2814566},
  note    = {Preprint: arXiv:1304.6475}
}

@article{bai2018greedy,
  author  = {Bai, Zhong-Zhi and Wu, Wen-Ting},
  title   = {On greedy randomized {Kaczmarz} method for solving large sparse linear systems},
  journal = {SIAM Journal on Scientific Computing},
  volume  = {40},
  number  = {1},
  pages   = {A592--A606},
  year    = {2018}
}

@article{bolukbacsi2024distributed,
  title={A distributed memory parallel randomized {Kaczmarz} for sparse system of equations},
  author={B{\"o}l{\"u}kba{\c{s}}{\i}, Ercan Sel{\c{c}}uk and Torun, Fahreddin {\c{S}}{\"u}kr{\"u} and Manguo{\u{g}}lu, Murat},
  journal={Concurrency and Computation: Practice and Experience},
  volume={36},
  number={25},
  pages={e8274},
  year={2024},
  publisher={Wiley Online Library}
}

@article{chazan1969chaotic,
  author  = {Chazan, Daniel and Miranker, Willard},
  title   = {Chaotic Relaxation},
  journal = {Linear Algebra and its Applications},
  volume  = {2},
  number  = {2},
  pages   = {199--222},
  year    = {1969},
  doi     = {10.1016/0024-3795(69)90028-7}
}

@article{coleman2026rwrj,
  title={Residual-Weighted Randomized {Jacobi}: Sharpened Bounds via Residual Concentration and Asynchronous Extension},
  author={Coleman, Evan},
  journal={arXiv preprint arXiv:2606.01232},
  year={2026}
}

@article{gordon1970algebraic,
  title={Algebraic reconstruction techniques ({ART}) for three-dimensional electron microscopy and {X}-ray photography},
  author={Gordon, Richard and Bender, Robert and Herman, Gabor T},
  journal={Journal of Theoretical Biology},
  volume={29},
  number={3},
  pages={471--481},
  year={1970},
  publisher={Elsevier}
}

@article{gordon2005carp,
  author  = {Gordon, Dan and Gordon, Rachel},
  title   = {Component-averaged row projections: a robust, block-parallel scheme for sparse linear systems},
  journal = {SIAM Journal on Scientific Computing},
  volume  = {27},
  number  = {3},
  pages   = {1092--1117},
  year    = {2005}
}

@article{gower2021adaptive,
  title={On adaptive sketch-and-project for solving linear systems},
  author={Gower, Robert M and Molitor, Denali and Moorman, Jacob and Needell, Deanna},
  journal={SIAM Journal on Matrix Analysis and Applications},
  volume={42},
  number={2},
  pages={954--989},
  year={2021},
  publisher={SIAM}
}

@article{haddock2019motzkin,
  title={On {Motzkin}'s method for inconsistent linear systems},
  author={Haddock, Jamie and Needell, Deanna},
  journal={BIT Numerical Mathematics},
  volume={59},
  number={2},
  pages={387--401},
  year={2019},
  publisher={Springer}
}

@article{haddock2021greed,
  title={Greed works: An improved analysis of sampling {Kaczmarz}--{Motzkin}},
  author={Haddock, Jamie and Ma, Anna},
  journal={SIAM Journal on Mathematics of Data Science},
  volume={3},
  number={1},
  pages={342--368},
  year={2021},
  publisher={SIAM}
}

@article{leventhal2010randomized,
  author  = {Leventhal, Dennis and Lewis, Adrian S.},
  title   = {Randomized methods for linear constraints: convergence rates and conditioning},
  journal = {Mathematics of Operations Research},
  volume  = {35},
  number  = {3},
  pages   = {641--654},
  year    = {2010}
}

@article{liu2014asynchronous,
  author  = {Liu, Ji and Wright, Stephen J. and Sridhar, Srikrishna},
  title   = {An asynchronous parallel randomized {Kaczmarz} algorithm},
  journal = {arXiv preprint arXiv:1401.4780},
  year    = {2014}
}

@article{moorman2021randomized,
  title={Randomized {Kaczmarz} with averaging},
  author={Moorman, Jacob D and Tu, Thomas K and Molitor, Denali and Needell, Deanna},
  journal={BIT Numerical Mathematics},
  volume={61},
  number={1},
  pages={337--359},
  year={2021},
  publisher={Springer}
}

@article{needell2015randomized,
  title={Randomized block {Kaczmarz} method with projection for solving least squares},
  author={Needell, Deanna and Zhao, Ran and Zouzias, Anastasios},
  journal={Linear Algebra and its Applications},
  volume={484},
  pages={322--343},
  year={2015},
  publisher={Elsevier}
}

@inproceedings{nutini2016convergence,
  author    = {Nutini, Julie and Sepehry, Behrooz and Laradji, Issam and Schmidt, Mark and Koepke, Hoyt and Virani, Alim},
  title     = {Convergence rates for greedy {Kaczmarz} algorithms, and faster randomized {Kaczmarz} rules using the orthogonality graph},
  booktitle = {Proceedings of the 32nd Conference on Uncertainty in Artificial Intelligence (UAI)},
  year      = {2016}
}

@article{panchal2026rgdbek,
  title={RGDBEK: Randomized Greedy Double Block Extended Kaczmarz Algorithm With Hybrid Parallel Implementation and Applications},
  author={Panchal, Aneesh and Behera, Ratikanta},
  journal={Numerical Linear Algebra with Applications},
  volume={33},
  number={3},
  pages={e70102},
  year={2026},
  publisher={Wiley Online Library}
}

@article{recht2011hogwild,
  title={{HOGWILD!}: A lock-free approach to parallelizing stochastic gradient descent},
  author={Recht, Benjamin and Re, Christopher and Wright, Stephen and Niu, Feng},
  journal={Advances in Neural Information Processing Systems},
  volume={24},
  year={2011}
}

@article{steinerberger2021weighted,
  title={A weighted randomized {Kaczmarz} method for solving linear systems},
  author={Steinerberger, Stefan},
  journal={Mathematics of Computation},
  volume={90},
  number={332},
  pages={2815--2826},
  year={2021}
}

@article{strohmer2009randomized,
  author  = {Strohmer, Thomas and Vershynin, Roman},
  title   = {A randomized {Kaczmarz} algorithm with exponential convergence},
  journal = {Journal of Fourier Analysis and Applications},
  volume  = {15},
  number  = {2},
  pages   = {262--278},
  year    = {2009}
}

@inproceedings{wang2022prkp,
  title={{PRKP}: A parallel randomized iterative algorithm for solving linear systems},
  author={Wang, Junjie and Tian, Min and Wang, Yinglong and He, Guoping and Liu, Tao},
  booktitle={2022 IEEE 24th Int Conf on High Performance Computing \& Communications; 8th Int Conf on Data Science \& Systems; 20th Int Conf on Smart City; 8th Int Conf on Dependability in Sensor, Cloud \& Big Data Systems \& Application (HPCC/DSS/SmartCity/DependSys)},
  pages={244--249},
  year={2022},
  organization={IEEE}
}

@article{zhang2023weighted,
  title={A weighted randomized sparse {Kaczmarz} method for solving linear systems},
  author={Zhang, Lu and Yuan, Ziyang and Wang, Hongxia and Zhang, Hui},
  journal={Computational and Applied Mathematics},
  volume={41},
  pages={383},
  year={2022},
  doi={10.1007/s40314-022-02105-9}
}

\end{document}